\documentclass[twocolumn, aps,pra]{revtex4-1}
\usepackage{graphicx}
\usepackage{epsfig}
\usepackage{color}

\begin{document}
\title{Ground-State Ferromagnetic Transition in Strongly Repulsive One-Dimensional Fermi Gases}
\author{Xiaoling Cui$^{\ast \dag}$ and Tin-Lun Ho$^{\ddag\ast}$}
\affiliation{$^{\ast}$ Institute for Advanced Study, Tsinghua University, Beijing 100084, China\\
$^{\dag}$Beijing National Laboratory for Condensed Matter Physics,
Institute of Physics, Chinese Academy of Sciences, Beijing 100190, China\\
$^{\ddag}$ Department of Physics, The Ohio State University, Columbus, OH 43210, USA}
\date{\today}

\begin{abstract}
We prove that as a one-dimensional Fermi gas is  brought across the resonance adiabatically from large repulsion to large attraction,  the singlet ground state will give way to the maximum spin state, which is the lowest energy state among the states accessible to the system in this process.  In the presence of tiny symmetry breaking fields that destroy spin conservation, the singlet ground state can evolve to the ferromagnetic state or a spin segregated state.  We have demonstrated  these effects by exact calculations on fermion cluster relevant to current experiments, and have worked out the quantum mechanical wavefunction that exhibits phase separation. 
\end{abstract}

\maketitle

Itinerant Ferromagnetism in metals is caused by interactions between electrons. The first theory of this phenomenon was given by Stoner 80 years ago using Hartree approximation\cite{Stoner}. Recently, the question of itinerant ferromagnetism has been investigated in the context of strongly repulsive three-dimensional(3D) Fermi gases\cite{ferro}. Despite an initial report of evidence of ferromagnetism\cite{Ketterle1} in Fermi gases, it was realized later that such phenomenon did not occur in these systems \cite{Ketterle2}.  For strongly repulsive Fermi gases, the issue of ferromagnetism is complicated by the severe atom loss due to three-body collisions\cite{Petrov}. This severe loss puts the system out of equilibrium, rendering the question of equilibrium ferromagnetism ill defined.

In a recent paper, we have pointed out through energy analysis\cite{CH} that the energy of the singlet state of a 1D homogeneous Fermi gas will  exceed that of the ferromagnetic state as the system crosses a resonance, implying  a transition from the singlet  to the ferromagnetic state. The important feature of 1D Fermi gas is that  it is  stable against atom loss near the transition\cite{Haller}. The problem of equilibrium ferromagnetism is therefore well defined.
Energetic consideration, however,  is not sufficient to guarantee experimental realization of  ferromagnetic transition. Due to the orthogonality of distinct spin wavefunctions of different ground states, the system may remain in the same spin state after crossing the transition point. This raises the question of how to facilitate and identify this transition.

Recently, Selim Jochim's group has succeeded in producing small clusters of Fermi gases in 1D harmonic traps\cite{Selim1, Selim2}.  Several theoretical studies have  found  ground state degeneracy in these systems in the strongly repulsive regime\cite{Zinner, Conduit, Lewenstein, Zinner2}. Some authors have attributed this to the tendency towards ferromagnetic\cite{Zinner,Conduit}.   With growing interests in itinerant ferromagnetism in large and small systems, it is useful to establish rigorous results for repulsive Fermi gases in arbitrary potentials. 

In this work, we shall establish a theorem on the level crossing between the singlet ground state  and the maximum spin state as the system passes through the resonance, where the coupling constant turns from $+\infty$ to $-\infty$.  Because of this crossing, a tiny perturbation that violates spin conservation can induce a dramatic transition from a singlet to a ferromagnetic state, or a segregation of spins.
The dramatic effects of symmetry breaking fields manifest themselves even down to small clusters. Through exact calculations, we show that even for a four fermion system, a tiny magnetic field gradient can cause a sudden spin segregation when the repulsion exceeds a critical value, 
a phenomena that can be observed in experiments similar to those in ref.\cite{Selim1, Selim2}. 
Our studies have also uncovered the exact wavefunction for the spin segregated state. As far as we know, this is the first explicit quantum mechanical wavefunction for a state with spins separated into magnetic domains.

Our paper is organized as follows. In section (I), we establish a theorem for ground state level crossing of the 1D repulsive Fermi gas across resonance. In section (II), we demonstrate the effect of symmetry breaking fields near the level crossing. In particular, we focus on the effect of a tiny magnetic field gradient, which induces a full spin segregation. The wave function for the spin segregated state and its many-body generalization will be presented in section (III). Finally the concluding remarks are made in section (IV).

\section{1D repulsive Fermi gases when swept across resonance} 

We  consider a two-component Fermi gas with $N$ particles with the Hamiltonian 
\begin{equation}
H=\sum_{i=1}^{N} h(x_{i}) + g\sum_{i>j}\delta (x_i - x_j),
\end{equation} 
where  
\begin{equation}
h(x) = -\frac{\hbar^2}{2M}\partial_{x}^2 + V(x),
\end{equation} 
$V(x)$ is a trapping potential, and $g\equiv \hbar^2 \gamma/M$ is the coupling constant.
The strong coupling regime corresponds to $\gamma/n\gg 1$, where $n$ is the density. 
 The two components will be referred to as $\uparrow$ and $\downarrow$ ``spin". In addition, we shall include a tiny transverse (radio-frequency) magnetic field 
 \begin{equation}
 V_h = h\sum_{i} \sigma_{i}^{x}, 
 \end{equation}
 and a tiny magnetic field gradient 
\begin{equation}
 V_{G} = -G\sum_{i} x_{i} \sigma_{i}^{z}.
\end{equation} 
$V_h$ and $V_G$ are extremely small perturbations that will not  cause any noticeable effects in a weakly interacting system. The effect of $V_h$ is to fix the direction of the ferromagnet, while $V_G$  destroys the spin conservation and  allows the system to reach  the lowest energy states within the family of states accessible to the system.

 Experimentally,   one can vary $1/g$ adiabatically from positive to negative through 0 by varying the external magnetic field\cite{Selim2}, a process we refer to as adiabatic sweep from the repulsive side, or simply ``adiabatic sweep". The point $1/g=0$ is referred to as the ``resonance"\cite{quasi1D}. 
The regions $1/g>0$ and $1/g<0$ will be referred to as the repulsive and the attractive side of the resonance. 
Near resonance on the attractive side,  $g\rightarrow -\infty$, the ground state  consists of  tightly bound pairs (molecules) with very large negative binding energy, $E_{b}= - Mg^2/(4\hbar^2)$ \cite{molecule}. In contrast, all fermions  on the repulsive side are in scattering states and therefore have positive energy. 
Due to the large energy difference between the scattering states and the tightly bound pairs, 
none of the states on the repulsive side can turn into the ground state on the attractive side in an adiabatic sweep. Instead, they all turn into excited  states with positive energies 
 on the attractive side.  These states will be referred to as super-Tonks states (in analogy to that in homogeneous systems\cite{Chen2}). {\it These are the only states accessible in the sweep process. }
From now on, when we talk about the lowest energy state after the adiabatic sweep, we mean the lowest energy state within the super-Tonks family.

\bigskip

We now consider the evolution of  the ground state of a Fermi gas in an adiabatic sweep.  We consider the cases with and without symmetry breaking fields separately. 

\bigskip

{\em Theorem on Level Crossing :  In an adiabatic sweep starting from the repulsive side of the resonance,  the energy of the singlet ground state
continues to increase, and will rise above  the energy  of  the maximum spin state, which is lowest energy state in the super-Tonks family. }
 
 \bigskip
 
{\em Proof.} Let  $E_{S}(-1/g)$ be the ground state energy of a repulsive Fermi gas with given total spin $S$ and interaction $g$. Its behavior across the resonance is determined by the following factors.

\bigskip

\noindent (A) According to Hellmann-Feynman theorem\cite{Feynman}, 
\begin{equation}
\frac{{\rm d}E}{{\rm }{d}(-1/g)}  \geq 0,
\end{equation} 
the ground state energy  $E_{S}(-1/g)$ increases with  $-1/g$\cite{continuous}. 
The equal sign holds for the maximum spin state $S=N/2$, whose energy is independent of interaction due to Pauli exclusion. 

\bigskip

\noindent (B) The Lieb-Mattis theorem\cite{Lieb} states that  for a repulsive Fermi gas  ($g>0$), the energy $E_{S}(-1/g)$ increases with total spin $S$, i.e. $E_{S_{1}} < E_{S_{2}}$ if $S_1<S_2$; 
and that the ground state of a Fermi gas with equal spin population is a singlet. 
This means that on the repulsive side of the resonance,  $E_S$ is bounded below by $E_{S=0}$ and above by $E_{S=N/2}$. (See Fig. 1). 

\bigskip

\noindent (C) 
At infinite repulsion ($1/g=0^{+}$), the ground states of all spin sectors are degenerate, i.e. $E_{S}(0) = E^{\ast}$ for all $S$. This result (which we shall prove momentarily) implies that close to resonance, 
  $E_{S}(-1/g) = E^{\ast} -\kappa_{S}/g$, where $\kappa_S = \left({\rm d}E_{S}(x)/{\rm d}x \right)_{x=0}$. 
  On the other hand, the ordering of $E_{S}$  for repulsive gas ($g>0$) as discussed in (B) implies that  
$\kappa_{S_1}>\kappa_{S_2}$ if $S_1<S_2$. Thus, on the attractive side, $g<0$, we have $E_{S_1}>E_{S_2}$ if $S_1<S_2$. The ordering of $E_S$ with spin on the attractive side is  just the opposite of that on the repulsive side, bounded below by $E_{S=N/2}$ and above by $E_{S=0}$.  (See Fig. 1).  
  This means that as the Fermi gas is swept through the resonance from the repulsive side, the energy $E_{S}(-1/g)$ will evolve as stated in the Theorem on level crossing. In other words,  the lowest energy gaseous state  will change from a singlet on the repulsive side to a maximum spin state on the attractive side.

\begin{figure}[hbtp]
\includegraphics[height=6cm,width=8.5cm]{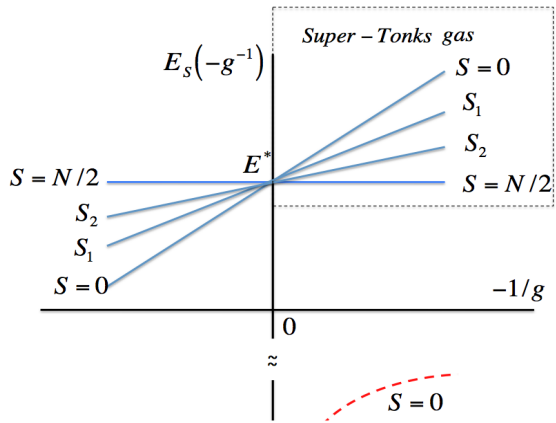}
\caption{ Schematic plot of the ground state energy $E_{S}(-1/g)$ of a Fermi gas with spin $S$ as a function of $-1/g$.  As $-1/g$ is swept across resonance adiabatically, each state with spin $S$ (solid line on the repulsive side $-1/g<0$) continuously evolves into a gaseous excited state in the super-Tonks family, rather than jumping to the true ground state consisting deep molecules (dashed line far below super-Tonks family). 
The positive slope of all $E_S$ is due to the Hellmann-Feynman theorem. On the repulsive side of the resonance,  $-1/g<0$,  Lieb-Mattis theorem requires that  $E_S$ increases with $S$, and the ground state is a singlet, ($N/2>S_2>S_1>0$).
The exact solution of the Schrodinger equation shows that $E_S$ must converge at the same point $E^{\ast}$ at resonance.  On the attractive side, $E_S$ increases with decreasing $S$. The super-Tonks family is bounded below by the maximum spin state $S=N/2$.  Thus, in  a  sweep process,  the lowest energy state will change abruptly from a singlet on the repulsive side to the maximum spin state on the attractive side. 
} \label{fig1}
\end{figure}

\bigskip

To prove (C), we note that at infinite interaction, the eigenstates of the system are of the form\cite{history}
\begin{eqnarray}
|\Psi\rangle = & \int D(1,2,.., N) \chi^{(S,S_z)}( 1,..N_{\downarrow}| N_{\downarrow}+1, .., N )  \nonumber  \\
     &  \times \prod_{i=1}^{N_{\downarrow}} 
     \psi^{\dagger}_{\downarrow}(i)  \prod_{j=N_{\downarrow}+1}^{N} \psi^{\dagger}_{\uparrow} (j) |0\rangle,
\label{Sm}  
 \end{eqnarray}
where $N_{\uparrow}+N_{\downarrow}=N$, $(1,2, ..)$ stand for the coordinates $(x_1, x_2, ..)$ of the fermions;   $\int$ means integrating over all $x_i$; 
$D(1, 2, ...N)$ is a Slater determinant of all the fermions (independent of spin) made up
of the eigenstates of $h(x)$, which is the wavefunction of a fully spin polarized state. The function $\chi^{(S,S_z)}( 1,..N_{\downarrow}| N_{\downarrow}+1, .., N )$ is  a spin eigenstate with total spin $S$ and magnetization $S_z =(N_{\uparrow}-N_{\downarrow})/2$, with down spins located at $ (x_{1} , ..x_{N_{\downarrow}}) $ and up spins at $(x_{N_{\downarrow}+1} , ..x_{N}) $.
This spin function is identical to that in the homogenous case, which is a constant in the regions  $(x_{P1} < x_{P2}< ... <x_{PN})$, where $\{ Pj \}$ stands for a permutation $P$ of the numbers $(1,2,..N)$\cite{Chen}\cite{1Dspinor}.  Eq.(\ref{Sm}) satisfies the Schrodinger equation in this region with energy given by that of a fully polarized state. QED. 

\bigskip

If the system has strict spin conservation, then despite the level crossing, the singlet ground state will remain a singlet on the attractive side and can not turn into the maximum spin state, ($S=N/2$). However, because of the large spin degeneracy at resonance, an  infinitesimal symmetry breaking field will have dramatic effects on ground state, as we shall show in Section (II) below. 

  \bigskip
  
Before proceeding, we would like to comment on the the process of adiabatic sweep. Since the spin excitations are all gapless at $1/g=0$, it seems that adiabaticity will be difficult to achieve near resonance.  As a practical matter, one can facilitate adiabaticity by turning on a small but non-zero field gradient so as to introduce a tiny energy scale. As long as the the sweep rate for $1/g$ is below this energy scale, the process will be adiabatic.  
Achieving adiabaticity is a general issue for all quantum phase transitions,  since the energies of the excitations vanish at the quantum critical point. To determine the presence of a quantum phase transition, strictly speaking, one will have to identity the quantum critical region through the expected scaling behavior in this region \cite{ZhouHo}.

\bigskip

\section{Effects of symmetry breaking fields near the level crossing }  

In the presence of a small transverse field $V_h$, the  ground state within each spin-$S$ sector is the state $|S, S_x = -S\rangle$ with maximum spin projection.  Still, $V_h$ conserves total  $S^2=(\sum_{i}{\bf s}_{i})^2$. The singlet ground state will remain a singlet on the attractive side after the sweep. The situation is changed completely if we further turn on the field gradient $V_G$.  Even if $V_{G}$  is much weaker than the (already small) perturbation $V_h$, the fact that it destroys the spin conservation means it will allow the singlet state to mix with other spin states and evolve to the ferromagnetic state $|S=N/2,S_x=-N/2\rangle$ after the adiabatic sweep. 
Another dramatic effect emerges when $V_h=0$ and $V_G\neq 0$.  Sufficiently close to resonance, $V_G$ will mix all the spin states to achieve a maximum reduction in energy. This results in a segregation into ferromagnetic domains, as we show below. 

\bigskip

 Consider four fermions (two $\uparrow$, two $\downarrow$)  in a harmonic trap with frequency $\omega_{T}$, and with a tiny field gradient $V_G$.
  We have  studied the ground state of this system by numerical exact diagonalization. The field gradient $G$ is chosen to be so small that it has no significant effects on the non-interacting Fermi gas,  as shown in Fig.2a. To be precise, the condition for such  ``insignificant" gradient is  
  \begin{equation}
  Ga_{ho}\ll \hbar \omega_{T},  \label{condition}
  \end{equation}
  where  $a_{ho}=\hbar/\sqrt{M\omega_{T}}$ is the length scale of the harmonic trap. . However, when the repulsion $g$  exceeds a critical value $g_{c}$, 
($g>g_{c}>0$ or $-1/g > -1/g_{c}$), 
 the spins suddenly separate into two domains on different sides of the system as shown in Fig.2b. This is the spin configuration that minimizes the Zeeman energy of $V_{G}$. 
For $g>g_c$, before reaching $g=+\infty$, the segregated spin density profile  quickly settles into that at $g=+\infty$ (or $1/g=0$),  indicating the quantum state  after $g>g_{c}$ quickly reaches that at $g=+\infty$.  Moreover, this spin segregated configuration remains unchanged after the system crosses the resonance on the attractive side where $g<0$.  Despite the dramatic change of spin density near resonance, we find that  the number density $n(x)= n_{\uparrow} +n_{\downarrow}$ remains unchanged, and is given by zero field gradient ($V_{G}=0$) configuration,  (Fig. 2c).  This  illustrates the huge difference in rigidity between charge and spin  at infinitely repulsion.  This formation of ferromagnetic domains is a direct consequence of the level crossing at $g=+\infty$ discussed in Section (I). The latter reflects the fact that the energies of spin excitations all vanish as $g\rightarrow +\infty$. Only when this happens can the spins in a singlet state be completely re-arranged into magnetic domains by a tiny field gradient. 

\bigskip

\begin{widetext}
 
\begin{figure}[hbtp]
\includegraphics[height=4.5cm,width=13cm]{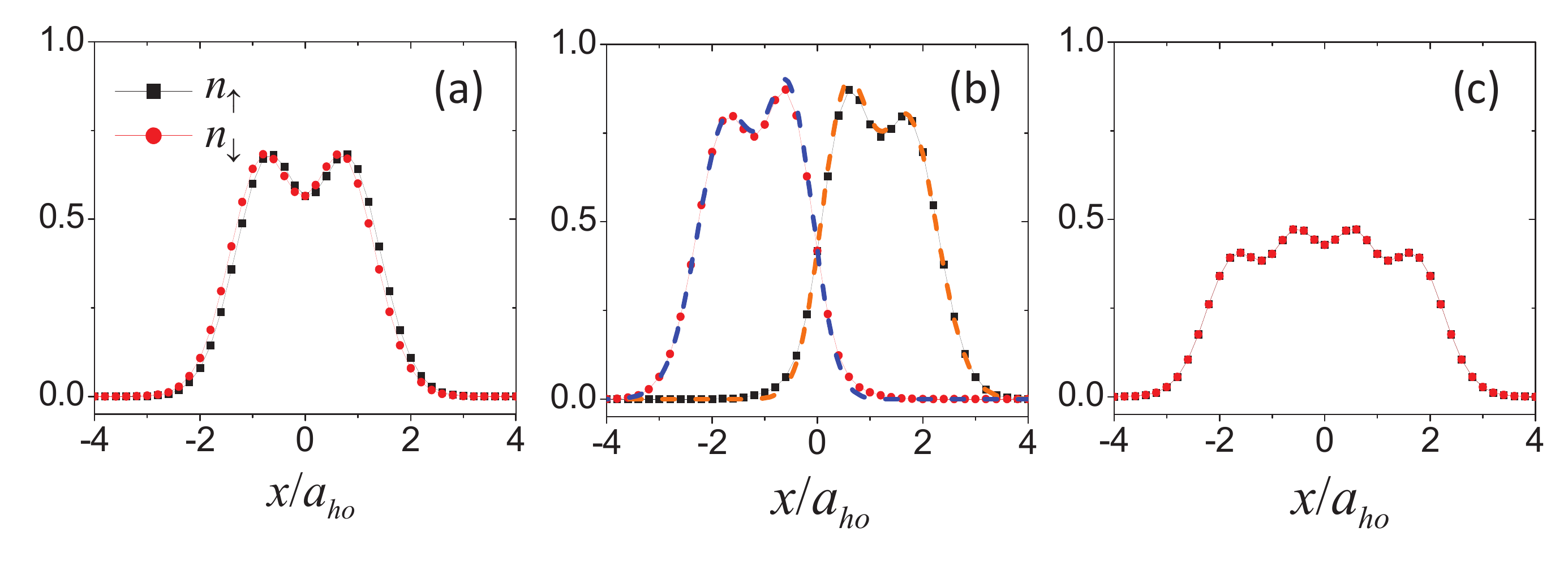}
\caption{(Color online) Density distributions for $\uparrow$(black square) and $\downarrow$(red circle) spins in a 
harmonic trap with frequency $\omega_{T}$ obtained from exact diagonalization. 
$N_{\uparrow}=N_{\downarrow}=2$. $a_{ho}=\hbar/\sqrt{M\omega_T}$. (a)$\gamma=0,\ G a_{ho}/\omega_{T}=0.05$;
(b)$\gamma a_{ho}=15,\ G a_{ho}/\omega_{T}=0.05$; (c) $\gamma a_{ho}=15,\ G=0$. 
In (b), the additional yellow (blue) dashed lines show the density of 
$\uparrow$ ($\downarrow$) spin obtained from performing degenerate perturbation on the six spin states at infinite repulsion (see text). The result matches well with that from exact diagonalization for large but finite repulsion. } \label{fig1}
\end{figure}

\end{widetext}


It is useful to look at this dramatic spin segregation in a different way.  If we start with zero field gradient ($G=0$) at  a large but finite repulsion, we find that  the ground state for the four fermion system is indeed a spin singlet as dictated by Lieb-Mattis theorem. In this case, the density profiles $n_{\uparrow}$ and $n_{\downarrow}$ for both spin components are identical as shown in Fig.2c.  However, as $G$ increases beyond a critical value $G_c$,
spin segregation suddenly occurs (Fig.2b), while the number density remain unchanged from $G=0$ case (Fig.2c). 

\bigskip

For the four-femion case,  we have found numerically that $G_{c}\rightarrow 0$  as $g^{-1} \rightarrow 0$. In fact, $G_c \sim 1/ g$ (for arbitrary particle number) as we shown below. That spin segregation can be activated suddenly by a vanishing small symmetry breaking field as the repulsion $1/g\rightarrow 0$ is a sign of phase transition. To see $G_c \sim 1/ g$, we compare the energies of the singlet ground state $|\Psi^{singlet}\rangle$ 
with that of the spin segregated state $|\Psi^{s-s}\rangle$  (both with equal spin population) in the presence of the field gradient $V_G$. For the singlet, 
$\langle V_G \rangle_{|\Psi^{singlet}\rangle }=0$, and its energy near resonance $1/g=0$ can be expanded as\cite{expansion, Chen2, Zwerger}, 
\begin{equation}
E^{singlet}(-1/g)= E^{\ast}\left(1- \frac{\alpha \hbar^2 n}{Mg} +  O(g^{-2})...\right),   
\end{equation}
where $E^{\ast}$ is the energy at $1/g=0$ and is the energy of a spin-polarized state. 
In contrast, the energy of the spin segregated state  $|\Psi^{s-s}\rangle$ is 
\begin{equation}
E^{s-s}= E^{\ast}- GX_{\uparrow} + GX_{\downarrow},
\end{equation}
where $X_{\uparrow, \downarrow} = \int {\rm d}x x n_{\uparrow, \downarrow}(x)$ is the center of mass of each magnetic domain.  Clearly, both $X_{\uparrow}$ and $X_{\downarrow}$ scales as  sample size $L$ and particle number $N$. 
Hence, we have  $(X_{\uparrow} - X_{\downarrow})=\beta NL $, where $\beta>0$ is a constant. The energy difference between these two states is then  
$ E^{singlet}-E^{s-s}  =  \beta NLG - E^{\ast} \alpha \hbar^2 n/(Mg)$. Spin segregation will occur if  
\begin{equation}
G> G_{c} = \frac{\alpha}{\beta} \frac{\hbar^2 n}{Mg}\frac{E^{\ast}}{N} \frac{1}{L}.
\end{equation} 
This shows that the critical $G_{c}$ vanishes with increasing repulsion $(g)$ and increasing sample size ($L$).  
Alternatively, for a given small $G$, the required interaction for the spin segregation is $g>g_c \propto G^{-1}$. The large value of $g_c$ (and thus small value of $1/g_c$) means the transition occurs very close to the resonance, and the spin configuration changes dramatically within a narrow region of $1/g$ with width $\delta(1/g)\propto G$.



\section{Wavefunction of the spin segregated state at infinite coupling} 

While we have obtained the spin segregated ground state (Fig.2b) for large  repulsion $g\rightarrow +\infty$   by exact diagonalization, the ground state at $g=+\infty$ can also be obtained from degenerate perturbation theory by diagonalizing $V_G$ within the space of degenerate spin states. 
This procedure is valid under Eq.(\ref{condition}) so that $V_G$ will not induce charge excitations but only re-organize the spin structure. 
Such a calculation will provide the explicit quantum mechanical wavefunction of a spin segregated state.

First, we note that in order to obtain the ground state under a tiny symmetry breaking field, it is essential to enumerate all the degenerate spin states $\{ \chi^{S, S_z}\} $ at infinite coupling. For fixed $N_{\uparrow}$ and $N_{\downarrow}$, the total number of degenerate spin states is $N!/(N_{\uparrow}!N_{\downarrow}!)$. Some of these states were previously obtained in Ref.\cite{Chen} by group theoretical method and in Ref.\cite{Girardeau} by generalized Fermi-Fermi mapping. However, the list there is incomplete. The complete set of these states can be obtained following the  procedure presented in Appendix A.

\begin{widetext}
For our four-particle case ($N_{\uparrow}=N_{\downarrow}=2$), there are six degenerate ground states at $g=+\infty$: 
 two $\chi^{(0,0)}$ states,  three  $\chi^{(1,0)}$ states, and  one  $\chi^{(2,0)}$ state. 
Following the general procedures in Appendix A, all the $|S,S_z=S\rangle$ states can be first constructed as
 \begin{eqnarray}
 \chi^{(2,2)}(1, 2, 3,4) &=& 1, \\
 \chi^{(1,1)}_{3} (1| 2, 3,4) &=&  [(1-P_{12}) + (1-P_{13}) + (1-P_{14})] [S(12)S(13)S(14)] ;    \label{chi113} \\ 
 \chi^{(1,1)}_{2} (1| 2, 3,4) &=& (1-P_{12})[S(12)(S(13)+S(14))] + (1-P_{13})[S(13)(S(12)+S(14) ) ]    \nonumber \\   
 && + (1-P_{14})[S(14)(S(12)+S(13) ) ] ;  \label{chi112} \\ 
 \chi^{(1,1)}_{1} (1| 2, 3,4) &=& (1-P_{12})S(12)+  (1-P_{13})S(13)+  (1-P_{14})S(14),   \label{chi111} \\
 \chi^{(0,0)}_{2} (1,2| 3,4)& = & [(1-P_{13})(1-P_{24}) + (1-P_{14}) (1-P_{23})] [S(13)S(14)S(23)S(24)] ;   \\
\chi^{(0,0)}_{1} (1,2| 3,4)& =&  (1-P_{13})(1-P_{24})[S(13)S(24)] + (1-P_{14}) (1-P_{23}) [S(14)S(23)]. 
\end{eqnarray}
where $P_{ij}$ denotes the permutation of the coordinates $x_i$ and $x_j$, and $S(ij)$ is the sign of  $(x_{i}-x_{j})$.  \\ \\
\end{widetext}

The states with $S_z=0$ can then be obtained by applying spin lower operator to above maximum $S_z(=S)$ states. After further normalization of each state, we finally get the explicit expressions of six degenerate wave functions as
\begin{eqnarray}
\chi^{(2,0)}&=&\frac{1}{\sqrt{6}};\nonumber\\
\chi_3^{(1,0)}&=&\frac{1}{\sqrt{8}}([1;234]+[2;134]);\nonumber\\
\chi_2^{(1,0)}&=&\frac{1}{\sqrt{24}}\Big( [1;234](S(13)+S(14)+2 S(12))\nonumber\\
 &&+ [2;134](S(23)+S(24)+2 S(21)) \nonumber\\
&&-[4;123](S(41)+S(42)+2 S(43))\nonumber\\
&&-[3;124](S(31)+S(32)+2 S(34)) \Big);\nonumber\\
 \chi_1^{(1,0)}&=&\frac{1}{\sqrt{40}}(S(14)+S(13)+S(24)+S(23));\nonumber\\ 
\chi_2^{(0,0)}&=&\frac{1}{\sqrt{48}}(3[12;34]-1);\nonumber\\
\chi_1^{(0,0)}&=&\frac{1}{4}(S(13)S(24)+S(14)S(23)).\nonumber
\end{eqnarray}
where $[12; 34] = S(13)S(14)S(23)S(24)$, $[1; 234]= S(12)S(13)S(14)$. \\

Given these expressions, it is straightforward to calculate the coupling induced by field gradient $V_G$ between all spin states, and then diagonalize the Hamiltonian in the degenerate manifold expanded by these states. Note that when constructing the H-matrix,  $\chi^{(1,0)}_{3}$ and $\chi^{(1,0)}_{1}$ need to be recombined as $(\chi^{(1,0)}_{3}\pm\chi^{(1,0)}_{1})/\sqrt{2}$, to ensure the orthogonality of these six states.
The resulted ground state is (over-bar means the state is normalized)
 \begin{eqnarray}
 |s-s\rangle =&  \sqrt{\frac{1}{6}} \,\, \overline{|\Psi^{(2,0)}_{}\rangle} +  \sqrt{\frac{1}{8}} \,\,  \overline{|\Psi^{(1,0)}_{3}\rangle} 
 - \sqrt{\frac{5}{8}} \,\,  \overline{|\Psi^{(1,0)}_{1}\rangle}+ \nonumber \\
 & \sqrt{\frac{1}{12}} \,\,  \overline{|\Psi^{(0,0)}_{2}\rangle}
+  \frac{1}{2} \,\,  \overline{|\Psi^{(0,0)}_{1}\rangle}, \hspace{0.5in}
\label{GG} \end{eqnarray}
which corresponds to the spin function 
\begin{eqnarray} 
 & & \chi^{s-s}(12 | 34) =  1 + [12; 34]+[1;234] +[2;134]  + S(13)S(24) \nonumber \\    &  &  \ \ \ \ \ \ \ +S(14)S(23) + S(41) + S(31) + S(42) + S(32) . \label{G}
 \end{eqnarray} 
 Evaluating Eq.(\ref{G}) in all regions $x_{P1}<x_{P2}<x_{P3}<x_{P4}$,  one finds that $\chi^{s-s}(12 | 34)$ is given by 
\begin{equation}
\chi^{s-s} (1,2| 1', 2') =\left\{\begin{array}{l} 1,\ \ \ {\rm if} \,\,\,\,\,  \{ 1, 2  \} < \{ 1', 2'\}; \\
0,\ \ \ {\rm otherwise}. \end{array}\right.
\label{SS} \end{equation}
where $ \{ 1, 2  \} < \{ 1', 2'\}$ means both $x_1$ and $x_2$ are to the left of $x_{1}'$ and $x_{2}'$. 
This is the spin function that leads to spin density in  Fig.2b.

By noting that the spin configuration in (\ref{SS}) exactly minimize the Zeeman energy of the field gradient for four particle system among all the configurations of lining-up spins, we can straightforwardly generalize this wave function to the many particle case as 
\begin{eqnarray}
&&\chi^{s-s} (1,2, .. , N_{\downarrow}| 1', 2', ..., N'_{\uparrow}) \nonumber\\
&=&\left\{\begin{array}{l} 1,\ \ \ {\rm if} \,\,\,\,\,  \{ 1, 2, .., N_{\downarrow}  \} < \{ 1', 2', ..., N'_{\uparrow} \}; \\
0,\ \ \ {\rm otherwise}. \end{array}\right.
\label{SS2} \end{eqnarray}
where $j'=N_{\downarrow}+j$, $N'_{\uparrow} = N_{\downarrow} + N_{\uparrow} = N$;  $i$ and $j'$ denote the coordinates $x_{i}$ and $x_{N_{\downarrow} + j}$ of down and up spins respectively. 
Eq.(\ref{SS2})  describes a spin segregated state for an arbitrary number of $\uparrow$ and $\downarrow$ spins. 
From (\ref{SS2}), we can estimate the location $(x_{o})$ of the domain wall, which is given by 
\begin{equation}
\int^{x_{o}}_{-\infty} {\rm d}x n(x)=N_{\downarrow}, 
\end{equation}
where $n(x)$ is the density determined by the determinant $D$ in Eq.(\ref{Sm}).  Note that above conclusions generally apply to any form of trapping potentials, as long as they are spin-independent.

To realize this spin segregation in alkali atoms, it is necessary to stay in  the low field regime where the magnetic moments of different spin states are different\cite{Paschen}. In the case of $^{40}K$,  there are spin states with Feshbach resonance near $B=200$G, and with magnetic moments  different enough to realize a sufficiently large field gradient term \cite{Kohl}. The presence of a Feshbach resonance allows one to tune the 3D  scattering length close to the  confinement length, which in turn allows one to achieve infinite 1D coupling through the confinement-induced resonance\cite{quasi1D}. For fermions that do not have a Feshbach resonance at low fields, as long as they have a positive scattering length, strong repulsion ( $\gamma/n\gg 1$) can be achieved by  making the system sufficiently dilute.

\section{ Concluding Remarks} 

The ferromagnetic transition discussed here is caused by the large spin degeneracy at infinite repulsion and the fact that the Fermi gas is locked in the excited branch when it passes through the resonance from the repulsive side.  While this is unique to 1D Fermi gases,  it is  remarkable that one can access a huge degenerate spin manifold by simply increasing repulsion. 
Our studies have demonstrated the dramatic effect of a tiny magnetic field gradient.  Other intricate ground states will surely emerge in the presence of  other types of symmetry breaking field (such as spin-orbit coupling) or residual spin-spin interaction. 
Strongly repulsive 1D Fermi Gases can be a very fertile ground for finding new types of spin ground states.

\bigskip

Notes Added: After the submission of our paper, a paper by S. E. Gharashi and D. Blume on a few particle cluster on harmonic trap has also appeared\cite{Blume}. The general setting and the role of symmetry breaking field discussed here were not considered there\cite{Blume}.  

\bigskip

We thank Randy Hulet for discussions. XC acknowledges the support of NSFC under Grant No. 11104158 and No. 11374177. TLH acknowledges the support by DARPA under the Army Research Office Grant Nos. W911NF-07-1-0464, W911NF0710576.

\bigskip

\bigskip

\noindent {\bf Appendix:  Construction of spin states $\chi^{(S,S_z)}$}

\bigskip

For a Fermi gas with 
$N_{\uparrow}$ up spins and $N_{\downarrow}$ down spins, the total number of degenerate spin states at infinite coupling is $N!/(N_{\uparrow}!N_{\downarrow}!)$.  All these states can be organized into different angular momentum states $\chi^{(S,S_z)}$, with $S_z=(N_{\uparrow}-N_{\downarrow})/2$ and $S=|S_z|,|S_z|+1,...,N/2$. In order to satisfy Fermi statistics, each $\chi^{(S,S_z)}$ is symmetric under the interchange of the coordinates of up spins and those of down spins respectively.  

\bigskip

To construct the spin state with given $(S,S_z)$, it is sufficient to  first construct the state with maximum spin projection $(S,S)$, 
since all other $(S,S_z)$ states can be obtained by applying the spin lowering operator $S_{-} = S_{x} - i S_{y}$ $P$-times on the state $(S,S)$, where $P=(S-S_z)$.  The general procedure to construct $(S,S)$ is :  


\bigskip

\noindent 
(i)  We start with the maximum spin state $\chi^{(N/2,N/2)}$ for a system of $N$ identical fermions,   $\chi^{(N/2,N/2)}=1$.

\noindent 
(ii) To construct the spin state $\chi^{(S,S)}$ with $S<N/2$, we contract an appropriate number of 
fermions pairs with opposite spins into singlets.  Because of Fermi statistics, interchanging the spins of two fermions amounts to interacting their coordinates times a minus sign. Hence, the state $\int D(1,2)(1-P_{12})S(12) \psi^{\dagger}_{\downarrow}(1) \psi^{\dagger}_{\uparrow}(2)|0\rangle$ represents a singlet pair of fermions, where $S(1,2)= {\rm sgn}(x_1-x_2)$.  The sign function $S(12)$ is needed because of the antisymmetry of $D(1,2)$. 
The fact that $S(12)$ is a constant (1 and -1)  in the regions $x_1 > x_2$ and $x_2> x_1$ means that product $D(1,2) S(1,2)$ satisfies the Schrodinger equation in these regions. 
To guarantee $S_z = S$,  the state has to be annihilated by $S_{+}= S_{x} +i S_{y}$ where ${\bf S}$ is the total spin of the system.

\noindent  (iii) After step (ii), one has to perform proper symmetrization so that the spin function is symmetric with respect to interchange of up spins and  interchange of down spins respectively. 

\bigskip

Note that in step (ii), there are multiple ways to contract singlet pairs. As  result, different maximum states $\chi^{(S,S)}$ with $S<N$ can be constructed. These $(S,S)$ states can be further orthogonalized to each other.  A straightforward analysis gives the number of orthogonal $(S,S)$ states as $(2S+1)N!/[(N/2+S+1)!(N/2-S)!]$.

\begin{widetext}

To illustrate this procedure, we consider the case $N=4$.  The wavefunction of the state  $\{ S=4, S_z = 4 \}$ is
\begin{equation}
| \Psi^{(S=4,S_z = 4)}_{3}\rangle = \int D(1,2,3,4)  \psi^{\dagger}_{\uparrow}(1) \psi^{\dagger}_{\uparrow}(2)  \psi^{\dagger}_{\uparrow}(3) \psi^{\dagger}_{\uparrow}(4) |0\rangle 
\end{equation}
To construct the spin states with $\{ S=3, S_z = 3 \}$, 
\begin{eqnarray} | \Psi^{(S=1,S_z = 1)}_{3}\rangle =&  \int D(1,2,3,4) \chi^{(1,1)}_{3} (1| 2, 3,4) \psi^{\dagger}_{\downarrow}(1) \psi^{\dagger}_{\uparrow}(2)  \psi^{\dagger}_{\uparrow}(3) \psi^{\dagger}_{\uparrow}(4) |0\rangle   \hspace{1.0in}   \label{11} 
\end{eqnarray} 
we first pick  two fermions to form a singlet pair. Let us say, we choose spin down fermion 1 and spin up fermion 2.  
This immediately have us the factor $(1-P_{12})S(12)$. 
Associate with this pair (12), the down spin $\psi^{\dagger}_{\downarrow}(1)$ can correlate with $j$ other spin up fermions with a sign function. For example, 
the factors  $(1-P_{12})[S(12)S(13)S(14)]$, $(1-P_{12})[S(12)S(13)]$, and $(1-P_{12})S(12)$ correspond to correlating the spin-down fermion (1) with 
3, 2, and 1 spin up fermion respectively.   Further symmetrizing among the up spins coordinates 
will then yield the spin wavefunctions shown in  Eq.(\ref{chi113}) to Eq.(\ref{chi111}). 
The eigenstate with $(S=1,S_z=0)$ is the 
\begin{eqnarray}  | \Psi^{(S=1,S_z = 0)}_{3}\rangle = &  \int D(1,2,3,4) [ \chi^{(1,1)}_{3} (1| 2, 3,4) +  \chi^{(1,1)}_{3} (2| 1, 3,4)]  \psi^{\dagger}_{\downarrow}(1) \psi^{\dagger}_{\downarrow}(2)  \psi^{\dagger}_{\uparrow}(3) \psi^{\dagger}_{\uparrow}(4) |0\rangle.  \label{10} \end{eqnarray} 

\end{widetext}

To show that $| \Psi^{(S=1,S_z = 1)}_{3}\rangle$ is a state of total spin $S=1$,  we note that it is annihilated by $S_{+}=S_{x}+iS_{y}$, which  then implies ${\bf S}^2 = S_{-}S_{+} + S_{z}^2 + S_{z}= S(S+1)=2$.  Noting that under the integral sign in Eq.(\ref{11}), the factor $1-P_{12}$ can be eliminated by turning $\psi^{\dagger}_{\downarrow}(1)\psi^{\dagger}_{\uparrow}(2)$ into a singlet pair $\psi^{\dagger}_{\downarrow}(1)\psi^{\dagger}_{\uparrow}(2)- \psi^{\dagger}_{\uparrow}(1)\psi^{\dagger}_{\downarrow}(2)$. This state  can therefore be annihilated by $S_{1+}+S_{2+}$. Since both 3 and 4 are spin up, the state is annihilated by total $S_{+}= \sum_{i=1}^{4} S_{i +}$.  



\bigskip

Following the same rule, we have written down two different singlet states for $N=4$ system, which are the functions  $\chi_{j}^{(0,0)}\ (j=2,1)$ in Eqs.(15,16) in the text.


\begin{thebibliography}{99}
\bibitem{continuous} The Hellmann-Feynaman theorem implicitly assumes that the energy is a continuous function of 1/g, including 1/g=0.  This continuity can be seen by noting that the g dependence of energy come from the behavior of the wavefunction when two particles approaches each other. For short-range interaction, such behavior is given by that of a two-particle system,  which can be shown to be continuous from the explicit wavefunction of the two-body solution. Similar continuity has been shown for the 3D case in the work by S. Tan,  Ann. Phys. {\bf 323}, 2971 (2008), and by S. Zhang and A. J. Leggett,  Phys. Rev. A {\bf 79}, 023601 (2009).
\bibitem{Paschen} In the high field regime,  all hyperfine states with the same electron spin have the same magnetic moment, and their energy differences will not be affected by the magnetic field gradient.
\end{thebibliography}
\end{document}